\documentclass{article}

\usepackage{microtype}
\usepackage{graphicx}
\usepackage{subfigure}
\usepackage{booktabs} 

\usepackage{hyperref}
\usepackage{url}

\usepackage[accepted]{icml2018}

\icmltitlerunning{DeepDrum: An Adaptive Conditional Neural Network}

\begin{document}

\twocolumn[
\icmltitle{DeepDrum: An Adaptive Conditional Neural Network\\
           for generating drum rhythms}



\icmlsetsymbol{equal}{*}

\begin{icmlauthorlist}
\icmlauthor{Dimos Makris}{equal,to}
\icmlauthor{Maximos Kaliakatsos-Papakostas}{equal,to2}
\icmlauthor{Katia Lida Kermanidis}{equal,to}

\end{icmlauthorlist}

\icmlaffiliation{to}{Department of Informatics, Ionian University, Corfu, Greece}
\icmlaffiliation{to2}{Institute for Language and Speech Processing, R.C. ``Athena'', Athens, Greece}

\icmlcorrespondingauthor{Dimos Makris}{c12makr@ionio.gr}


\icmlkeywords{LSTM, Neural Networks, Deep learning, Rhythm Composition}

\vskip 0.3in
]

\printAffiliationsAndNotice{}

\begin{abstract}
Considering music as a sequence of events with multiple complex dependencies, the Long Short-Term Memory (LSTM) architecture has proven very efficient in learning and reproducing musical styles. However, the generation of rhythms requires additional information regarding musical structure and accompanying instruments. In this paper we present DeepDrum, an adaptive Neural Network capable of generating drum rhythms under constraints imposed by Feed-Forward (Conditional) Layers which contain musical parameters along with given instrumentation information (e.g.\ bass and guitar notes). Results on generated drum sequences are presented indicating that DeepDrum is effective in producing rhythms that resemble the learned style, while at the same time conforming to given constraints that were unknown during the training process.
\end{abstract}

\section{Introduction}
Developing computational systems that can be characterized as creative \cite{deliege2006musical} has long been the focus of research. Such systems raging tasks from melody and chords' composition to lyrics and rhythm. With the advent of Deep Learning architectures numerous research works have been published using different types of Artificial Neural Networks and, especially the Long Short-Term Memory (LSTM) networks for composing music since these are capable of modeling sequences of events (e.g. \cite{hadjeres2016deepbach,kalingeri2016music} and \cite{briot2017deep} for further references).

To the best of our knowledge there is limited word addressing learning and generating percussion (drum) progressions with such architectures \cite{hutchings2017talking,choi2016text}. Most of these methods focus on the generation of sequences that imitate a learned style. However, the performance of human drummers is potentially influenced, e.g.\ by what the guitar and bass player plays, while the tempo of a song affects the density of their playing. 

In this work we present DeepDrum, a combination of LSTMs and Feed-Forward (FF), or Conditional, modules capable of composing drum sequences based on musical parameters, i.e. guitar and bass performance, tempo, grouping and metrical information. DeepDrum is able to combine implicitly learned information (LSTM) and explicitly defined conditions (FF), allowing the generation of drum rhythms that resemble a musical styles through implicit learning and, at the same time, satisfy some explicitly declared conditions that are potentially not encountered in the learned style.

\section{Data Collection and Architecture Information} 
The utilised corpus consists of 70 songs from two progressive rock bands, that have common musical characteristics, and were collected manually from web tablature learning sources \footnote{ \url{http://www.911tabs.com/}}. Following the same methodology for conditional composition, similarly to \cite{makris2017combining}, the representation of input training data was based on text words with one-hot encodings. 

The proposed architecture comprises separate modules for predicting the next drum event. The LSTM module learns sequences of consecutive drum events, while the Conditional (FF) module handle musical information regarding guitar, bass, metrical structure, tempo and grouping. This information is the sum of features of consecutive time-steps, in one-hot encodings, of the Conditional Input space within a moving \textit{window} giving information about the past, current and future time-steps. 

\begin{figure}[t]
\includegraphics[width=0.45\textwidth]{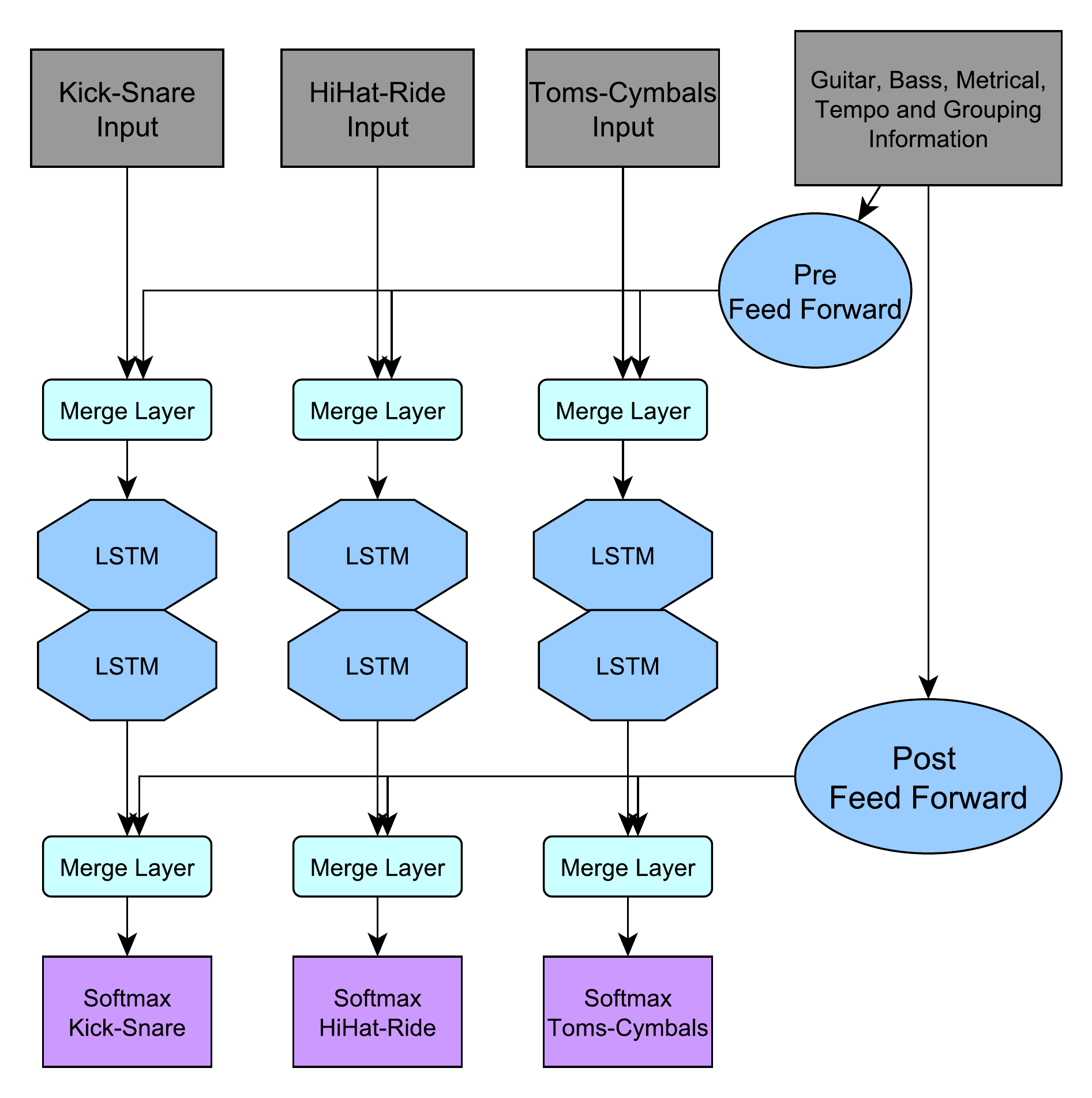}
\caption{DeepDrum Neural Network Architecture.}
\centering
\label{fig:NN_arch}
\end{figure}

DeepDrum has 3 input spaces for different elements of drums, thus leading to 3 LSTM Block modules, while the Conditional input space is separated to 2 FF modules. The Pre-FF carries information for the past time-steps and is merged with each corresponding drum input. The Post-FF contains information for current and future time-steps which are merged with each LSTM block output, thus leading to independent softmax outputs. Concerning the configuration, two stacked LSTM layers and single FF layers (linear activation) with 256 Hidden Units were used along with dropouts of $0.2$ in every connection. In our experiments we used Keras \cite{chollet2015keras} library with Tensorflow \cite{abadi2016tensorflow} deep learning framework as backend. Figure \ref{fig:NN_arch} summarises the proposed architecture.

\section{Experimental Setup} \label{sec:results}

The introduced architecture is examined for producing drums rhythms according to the conditions given by pieces that were not included in the training set, with some of them having musical characteristics that have been not encountered in any piece of the training corpus (e.g.\ time signatures 3/8, 9/8). These four pieces, however, pertain to the learned style of the training corpus (denoted as $PT$-$PF$). In addition we used 2 pieces from a different genre (in Disco style denoted as $AB$). Multiple generations were produced using initial seed-sentences, in different stages of the learning process with adjustable diversity parameter. Interested readers can listen to several excerpts generated with the proposed architecture on the web page of Humanistic and Social Informatics Lab of Ionian University \footnote{https://hilab.di.ionio.gr/index.php/en/deepdrum-an-adaptive-conditional-neural-network-for-generating-drum-rhythms/}. 

\begin{figure}[t]
\includegraphics[width=0.48\textwidth]{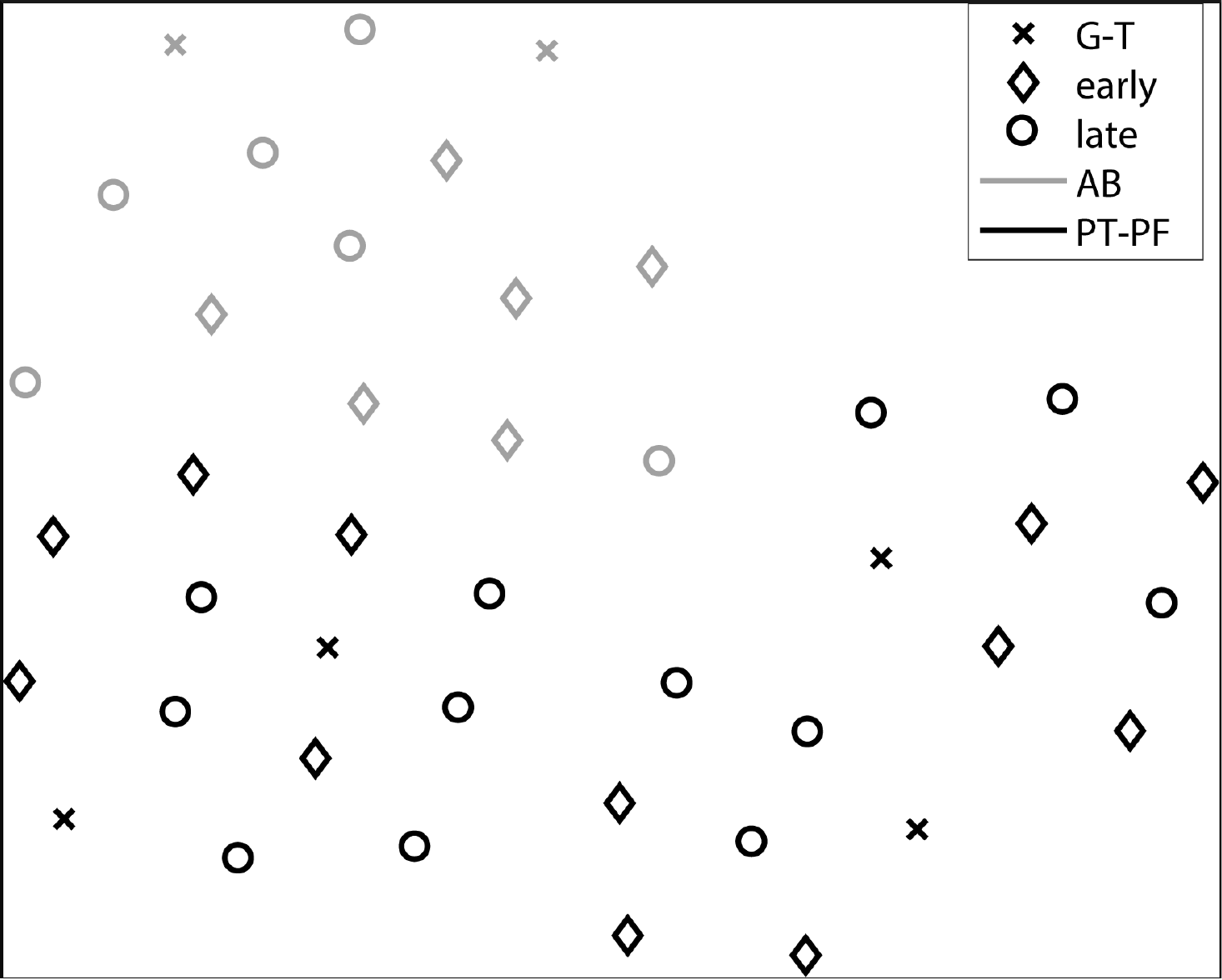}
\caption{Two-dimensional mapping of the features of all DeepDrum compositions. Features of the Ground-Truth (G-T) rhythms are illustrated with a $\times$ symbols, while the features of rhythms with less (early - $\circ$) and more (late - $\diamond$) than 100 epochs are shown separately.}
\centering
\label{fig:rnn_ff_all}
\end{figure}

Drum rhythm features~\cite{kaliakatsos2013evodrummer} were extracted from the generated pieces. The mean values and standard deviations of these features covering all the bars of the generated content were used as global features of each piece. Figure~\ref{fig:rnn_ff_all} illustrates the two dimensional reduction of the global features of all pieces using the t-SNE~\cite{van2009learning} technique. We can notice that: a) the network composes $AB$ pieces that approach the features of this style (late generations are closer) while b) $PT$-$PF$ pieces composed with unknown conditions of the learned styles cover areas around the corresponding Ground-Truth pieces.

\vspace{-5pt}
\section{Conclusions} \label{sec:conclusions}
This work introduces DeepDrum, an adaptive Neural Network application which learns and generates sequences under given musical constraints. The proposed architecture consists of a Recurrent module with LSTM blocks that learns sequences of consecutive drums' events along with two Feed-Forward (Conditional) Layers handling information for musical instruments, metrical structure, tempo and the grouping (phrasing).

The results shows the importance of the Conditional Layers which enable DeepDrum to simulate humans drummers in two tasks: respond to create ``groove" with other instruments in any musical style, and foresee future musical changes (e.g.\ phrase and tempo changes). In addition, the Conditional Layers allows to keep the entire network ``on-track'' and enable it to respond to constraints that were not encountered during training (e.g.\ unknown -- to the network -- time signatures).

\section*{Acknowledgements}
This research has been financially supported by General Secretariat for Research and Technology (GSRT) and the Hellenic Foundation for Research and Innovation (HFRI) (Scholarship Code: 953).

\bibliography{icml}

\begin{thebibliography}{11}
\providecommand{\natexlab}[1]{#1}
\providecommand{\url}[1]{\texttt{#1}}
\expandafter\ifx\csname urlstyle\endcsname\relax
  \providecommand{\doi}[1]{doi: #1}\else
  \providecommand{\doi}{doi: \begingroup \urlstyle{rm}\Url}\fi

\bibitem[Abadi et~al.(2016)Abadi, Barham, Chen, Chen, Davis, Dean, Devin,
  Ghemawat, Irving, Isard, et~al.]{abadi2016tensorflow}
Abadi, Mart{\'\i}n, Barham, Paul, Chen, Jianmin, Chen, Zhifeng, Davis, Andy,
  Dean, Jeffrey, Devin, Matthieu, Ghemawat, Sanjay, Irving, Geoffrey, Isard,
  Michael, et~al.
\newblock Tensorflow: A system for large-scale machine learning.
\newblock In \emph{OSDI}, volume~16, pp.\  265--283, 2016.

\bibitem[Briot et~al.(2017)Briot, Hadjeres, and Pachet]{briot2017deep}
Briot, Jean-Pierre, Hadjeres, Ga{\"e}tan, and Pachet, Fran{\c{c}}ois.
\newblock Deep learning techniques for music generation-a survey.
\newblock \emph{arXiv preprint arXiv:1709.01620}, 2017.

\bibitem[Choi et~al.(2016)Choi, Fazekas, and Sandler]{choi2016text}
Choi, Keunwoo, Fazekas, George, and Sandler, Mark.
\newblock Text-based lstm networks for automatic music composition.
\newblock \emph{arXiv preprint arXiv:1604.05358}, 2016.

\bibitem[Chollet et~al.(2015)]{chollet2015keras}
Chollet, Fran{\c{c}}ois et~al.
\newblock Keras, 2015.

\bibitem[Deli{\`e}ge \& Wiggins(2006)Deli{\`e}ge and
  Wiggins]{deliege2006musical}
Deli{\`e}ge, Ir{\`e}ne and Wiggins, Geraint~A.
\newblock \emph{Musical creativity: Multidisciplinary research in theory and
  practice}.
\newblock Psychology Press, 2006.

\bibitem[Hadjeres \& Pachet(2016)Hadjeres and Pachet]{hadjeres2016deepbach}
Hadjeres, Ga{\"e}tan and Pachet, Fran{\c{c}}ois.
\newblock Deepbach: a steerable model for bach chorales generation.
\newblock \emph{arXiv preprint arXiv:1612.01010}, 2016.

\bibitem[Hutchings(2017)]{hutchings2017talking}
Hutchings, P.
\newblock Talking drums: Generating drum grooves with neural networks.
\newblock \emph{arXiv preprint arXiv:1706.09558}, 2017.

\bibitem[Kaliakatsos-Papakostas et~al.(2013)Kaliakatsos-Papakostas, Floros, and
  Vrahatis]{kaliakatsos2013evodrummer}
Kaliakatsos-Papakostas, M., Floros, A., and Vrahatis, M.~N.
\newblock Evodrummer: deriving rhythmic patterns through interactive genetic
  algorithms.
\newblock In \emph{International Conference on Evolutionary and Biologically
  Inspired Music and Art}, pp.\  25--36. Springer, 2013.

\bibitem[Kalingeri \& Grandhe(2016)Kalingeri and Grandhe]{kalingeri2016music}
Kalingeri, Vasanth and Grandhe, Srikanth.
\newblock Music generation with deep learning.
\newblock \emph{arXiv preprint arXiv:1612.04928}, 2016.

\bibitem[Makris et~al.(2017)Makris, Kaliakatsos-Papakostas, Karydis, and
  Kermanidis]{makris2017combining}
Makris, Dimos, Kaliakatsos-Papakostas, Maximos, Karydis, Ioannis, and
  Kermanidis, Katia~Lida.
\newblock Combining lstm and feed forward neural networks for conditional
  rhythm composition.
\newblock In \emph{International Conference on Engineering Applications of
  Neural Networks}, pp.\  570--582. Springer, 2017.

\bibitem[van~der Maaten(2009)]{van2009learning}
van~der Maaten, Laurens.
\newblock Learning a parametric embedding by preserving local structure.
\newblock \emph{RBM}, 500\penalty0 (500):\penalty0 26, 2009.

\end{thebibliography}
\bibliographystyle{icml2018}

\end{document}